%% file: A_A_Draft.tex
\documentclass{aa}

\usepackage{amsmath,amsfonts,amssymb,xcolor}
\usepackage{graphicx}
\usepackage[varg]{txfonts} 
\usepackage[english]{babel}
\usepackage{natbib} 
\usepackage{url}
\usepackage{mathtools} 
\usepackage{sansmathaccent}
\usepackage{gensymb}
\usepackage{bm}

\bibpunct{(}{)}{;}{a}{}{,} 

\definecolor{darkgreen}{RGB}{0,100,0}
\definecolor{darkblue}{RGB}{0,0,139}

\newcommand{\GG}[1]{}

\begin{document}

   \title{Element history of the Laplace resonance: a dynamical approach}


   \author{F. Paita\inst{1}
          \and
          A. Celletti\inst{2}
                                        \and
                                        G. Pucacco\inst{3}
          }

   \institute{Department of Mathematics, University of Rome Tor Vergata,
              Via della Ricerca Scientifica 1, 00133 Rome\\
              \email{f.paita@live.it}
         \and
                                                        Department of Mathematics, University of Rome Tor Vergata,
              Via della Ricerca Scientifica 1, 00133 Rome\\
              \email{celletti@mat.uniroma2.it}
                                 \and
                Department of Physics, University of Rome Tor Vergata,
                                                        Via della Ricerca Scientifica 1, 00133 Rome\\
                \email{pucacco@roma2.infn.it}
             }

   \date{Received <date> \
          Accepted <date>}


  \abstract
   {We consider the three-body mean motion resonance defined by the Jovian moons Io, Europa, and Ganymede, which is commonly known as the Laplace resonance. In terms of the moons' mean longitudes $\lambda_1$ (Io), $\lambda_2$ (Europa), and $\lambda_3$ (Ganymede), this resonance is described by the librating argument $\phi_L \equiv \lambda_1-3\lambda_2+2\lambda_3\approx 180\degree$, which is the sum of $\phi_{12} \equiv\lambda_1-2\lambda_2+\varpi_2 \approx 180\degree$ and $\phi_{23} \equiv \lambda_2-2\lambda_3+\varpi_2 \approx 0\degree$, where $\varpi_2$ denotes Europa's longitude of perijove.}
   {In particular, we construct approximate models for the evolution of the librating argument $\phi_L$ over the period of 100 years, focusing on its principal amplitude and frequency, and on the observed mean motion combinations $n_1-2n_2$ and $n_2-2n_3$ associated with the quasi-resonant interactions above.}
        {First, we numerically propagated the Cartesian equations of motion of the Jovian system for the period under examination, and by comparing the results with a suitable set of ephemerides, we derived the main dynamical effects on the target quantities. Using these effects, we built an alternative Hamiltonian formulation and used the normal forms theory to precisely locate the resonance and to semi-analytically compute its main amplitude and frequency.}
   {From the Cartesian model we observe that on the timescale considered and with ephemerides as initial conditions, both $\phi_L$ and the diagnostics $n_1-2n_2$ and $n_2-2n_3$ are well approximated by considering the mutual gravitational interactions of Jupiter and the Galilean moons (including Callisto), and the effect of Jupiter's $J_2$ harmonic. Under the same initial conditions, the Hamiltonian formulation in which Callisto and $J_2$ are reduced to their secular contributions achieves larger errors for the quantities above, particularly for $\phi_L$. By introducing appropriate resonant variables, we show that these errors can be reduced by moving in a certain action-angle phase plane, which in turn implies the necessity of a tradeoff in the selection of the initial conditions.}
   {In addition to being a good starting point for a deeper understanding of the Laplace resonance, the models and methods described are easily generalizable to different types of multi-body mean motion resonances. Thus, they are also prime tools for studying the dynamics of extrasolar systems.}


   \maketitle

%
%

\section{Introduction}

The Jovian system holds a special place in planetary science at least since the time of Galileo, and Jupiter itself has been tracked by Babylonian and Chinese astronomers. Contemporary dynamical theories of the system have shown that at least for the Galilean moons, the major perturbations of the orbits arise from the oblateness of Jupiter and the mutual gravitational interactions among the satellites, thus making them part of a miniature solar system. Furthermore, in the last years of space exploration, several missions (Galileo, Juno, and Juice) have been sent or will be launched to refine our understanding of the system's physical characteristics, and how they relate to the dynamics of the system.

\begin{figure}[t]
\resizebox{\hsize}{!}{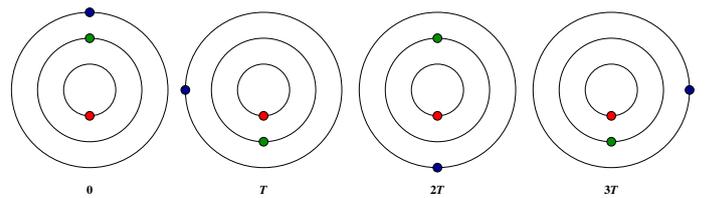}
\caption{Jovicentric snapshots of the Jovian system, with $T$ denoting Io's mean orbital period. The moons are color-coded (Io in red, Europa in green, and Ganymede in blue).}
\label{bicGeo}
\end{figure}
The key dynamical aspect at the core of these studies is the so-called Laplace resonance, a three-body mean motion resonance coupling the dynamics of the moons Io, Europa, and Ganymede. This resonance, whose main geometrical effect is the prevention of a triple conjunction of the moons (see Figure \ref{bicGeo}), is the product of two different 2:1 quasi-resonant interactions, which can be described in terms of the observed mean motions ($n_1$ for Io, $n_2$ for Europa, and $n_3$ for Ganymede) as \citep{Br}

\begin{equation}
\begin{aligned}
n_1 - 2n_2 = 0.7395507361\ \degree / day,\\
n_2 - 2n_3 = 0.7395507301\ \degree / day.
\end{aligned}
\label{laplMM}
\end{equation}
In addition to these two, a weaker resonant interaction exists between Ganymede and Callisto, quantifiable as $3n_3-7n_4 = -0.049084320\ \degree / day$ (again, see \citet{Br}).

The relations in Eq. \eqref{laplMM} were first shown by Laplace \citep{Lapl} to imply the existence of the librating angles

\begin{equation}
\begin{aligned}
\lambda_1 -2\lambda_2 + \varpi_2 &\approx 180\degree,\\
\lambda_2 -2\lambda_3 + \varpi_2 &\approx 0\degree,
\end{aligned}
\end{equation}
where $\lambda_i$ denotes the mean longitudes of the moons and $\varpi_i$ the longitudes of the perijoves. These equations combine in the Laplace argument $\phi_L\equiv \lambda_1 - 3\lambda_2 + 2\lambda_3\approx 180\degree$. Furthermore, because $\varpi_1$ and $\varpi_2$ drift with similar rates \citep{YoPe}, the system is affected by the additional librating relation $\lambda_1 -2\lambda_2 + \varpi_1 \approx 0\degree$.

Laplace's theory was the first that could be considered relatively complete, but it was not specifically aimed to generate ephemerides for the system. Another model, similar in scope and important for our work, was developed some decades later by de Sitter \citep{Si}. It uses intermediary orbits, which as shown in \citet{BrHa} are periodic solutions of the differential equations obtained by retaining exclusively resonant terms. Very recently, we have shown \citep{CePaPu} that the stable configuration of this system differs from the actual configuration because the latter possesses a rotating angular combination that is fixed in the former.

At about the same time as de Sitter, Sampson \citep{Sa} developed his own theory on the motion of the Galilean moons, tailored to easily generate ephemerides tables. This was later corrected and expanded by Lieske \citep{Li-B,Li-A}, whose E5 ephemerides are the reference today for precisely tracking the orbits of the Galilean moons, as reported, for example, by \citet{MuVaMoSc}, \citet{La-A}, and \citet{Ko}. 

In parallel to these ``astronomical'' works, the search for explicit approximate solutions of the system saw the introduction of additional theories, even Hamiltonian ones, starting from the 1970s. First in chronological order, Marsden \citep{Marsd} applied von Zeipel's method \citep{Mo} in his PhD thesis to average out the short-period terms, and he then solved the resulting differential equations for the long-period effects by successive substitutions. Later on, Sagnier \citep{Sag} and Ferraz-Mello \citep{Fe,FeBook} worked with complex variables and Taylor expansions to find their solutions, particularly assuming fixed the resonant frequencies. Different sets of variables were used by Brown \citep{Br}, who exploited a Lie transform normalization \citep{Ka,GLS} to remove short-period terms in his non-Hamiltonian formulation. The latter was instead employed by Henrard \citep{He}, who reduced his system to 4 d.o.f. (hereafter, short for degrees of freedom) and then explicitly computed amplitude and frequency of the Laplace resonance using a similar technique.

We here continue several of these works in the sense that our overall objective is the construction of approximate models for the study of the Laplace resonance over short timescales and not necessarily precise ephemerides of the Galilean moons (as done, e.g., in \citet{CePaPu}). Similarly to \citet{CePaPu-B} and \citet{Lari}, the goal is extrapolating information on the dynamics of this particular resonance and provide a baseline reference for future analyses of its mechanisms.

By eschewing the search for precise ephemerides, we are able to obtain several advantages. In particular, we present models that use the fewest parameters possible to describe the Laplace resonance with a sufficient degree of precision, that is, the values for the main amplitude and frequency of the libration mostly agree with those reported in the literature \citep{Li-B, MuVaMoSc}. Thanks to their simplicity, they are easy to reproduce and can be reused in conjunction with different objectives and approaches. The caveat, and this is an important point, is that they show a strong sensitivity to the initial conditions, especially with regard to the amplitude of $\phi_L$. Still, we show that by exploiting the inherent dynamics of the system and normal form techniques, it is possible to construct a procedure to recover what we need, as long as a tradeoff on the initial conditions is taken into account.

The first step is to introduce a benchmark model for the amplitude and frequency of the resonant argument. While precise ephemerides are not a concern, we have found Lainey's Cartesian formulation \citep{LaThesis} to be particularly suitable for our purposes. Thus, after reviewing his model and trimming it to the fundamental dynamical contributions, we extend his work by evaluating the evolution of the Laplace argument over the period of 100 years, comparing it to NASA's ephemerides, and extracting the fundamental amplitude and frequency that we consider in the rest of the paper.

The main model of this work is Hamiltonian, which suits our analytical objectives because of the powerful results associated with this formulation. The basic, planar version of the model presented here rests on a chain of suitable series expansions and the retaining of terms corresponding to the 2:1 quasi resonant interactions described at the beginning (plus the secular effect of Jupiter's $J_2$ harmonic). This degree of development is sufficient to analyze the validity of the model, and through numerical comparison with the Cartesian formulation, we show that under the same initial conditions, the Hamiltonian adequately approximates the Galilean moons' dynamics in terms of eccentricities and semi-major axes. Amplitude and frequency of the Laplace argument, as well as the diagnostics $n_1-2n_2$ and $n_2-2n_3$, are instead more difficult to recover. In particular, under a suitable set of canonical coordinates, we apply the normalization procedure previously mentioned to focus on a certain equilibrium of the system, obtain a 1 d.o.f. normal form Hamiltonian, and exploit its characteristics to approximate the quantities we search for.

The paper is organized as follows. In Section 2 we present the Cartesian formulation employed as baseline for the history of the Laplace argument, and we compare it numerically with a set of ephemerides in order to justify the dynamical effects included in the model. The conclusions of this section are then exploited in Section 3 to construct the alternative, planar Hamiltonian formulation for the Laplace resonance. The validity of this is subsequently confirmed in Section 4, where a numerical comparison is drawn with the results associated with the Cartesian model and where we show how appropriate values for the desired quantities can be recovered. Finally, in Section 5 we summarize the main results of this work and present parallel and future directions of research.

%
%

\section{Benchmark model for the Laplace argument}


\subsection{Introduction and reference frames}

In this section we introduce a Cartesian model that  precisely captures the evolution of the Laplace argument (over the period of 100 years). Through straightforward numerical integrations, we compare the results of this model with the equivalent results associated with a given set of ephemerides, and we exploit these comparisons to single out the dynamical effects that are used in the successive Hamiltonian formulation. Key variables for these comparisons are not only amplitude and period of the Laplace argument, but because of the composite nature of the resonance, also the 2:1 quasi-resonant observed mean motion combinations $n_1-2n_2$ and $n_2-2n_3$.

The equations of motion are defined and integrated in a Jovicentric fixed frame. Here we used the ``J2000'' frame implemented in the NASA Spice toolkit\footnote{\url{https://naif.jpl.nasa.gov/naif/toolkit.html}}, from which we also extracted the set of ephemerides giving the history of the Laplace argument. As indicated in the Spice documentation, the previous frame is assumed to be individualized by Earth's mean equatorial plane at the J2000 epoch (positive $x$ -axis parallel to the vernal equinox direction) and the corresponding normal (with Earth's spin axis indicating the positive $z$ direction). In reality, it corresponds to the so-called International Celestial Reference Frame, but the two are separated by a rotation of just $0.1$ arcsecond. For more detail, we refer to the Spice documentation\footnote{\url{https://naif.jpl.nasa.gov/pub/naif/toolkit_docs/Tutorials/pdf/individual_docs/17_frames_and_coordinate_systems.pdf}}.

The angular librating relations characterizing the Jovian system depend on the osculating orbital elements, which in turn are defined through Jovicentric \textup{equatorial} coordinates. Thus, before visualization, the state vectors are transformed from the J2000 frame into a frame defined by Jupiter's mean equatorial plane and the corresponding spin axis. The latter defines the $z$ -axis of this frame, while the line of nodes obtained by intersecting the previous equatorial plane with the J2000 plane acts as the associated $x$ -direction. We note that this new frame is non-inertial, since Jupiter's spin axis is precessing (we do not take nutation into account). However, since this effect is very small, we considered it only for the representation of the ephemerides, while for the integrated flow, we froze the rotation at the initial integration epoch. A complete graphical visualization of the two frames (J2000 and mean equatorial) is given in Figure \ref{Prc_Frame}.

This section is organized as follows. In Section \ref{MotEqsSec} we introduce our notations and derive the general form of the equations of motion. Following this, in Section \ref{PrecSec} we discuss the precession of Jupiter's mean north pole, which, as shown in Section \ref{OblSec}, plays a role in how the associated oblateness potential is shaped. In Section \ref{NumConsSec} we provide some technical details on the numerical simulations we performed, whose results are then described in Section \ref{FigSec}.

To conclude, we observe that much of the content of this section, along with the notation we adopted, follows \citet{La-A} and \citet{LaThesis}. The focus of these works is the construction of an accurate set of ephemerides for the Galilean moons. Here instead we determine reference values and characteristics for the successive Hamiltonian formulation.


\subsection{Equations of motion}\label{MotEqsSec}

The first step for constructing the Cartesian formulation of the Galilean moons' dynamics is to consider the system they form with Jupiter as isolated. These bodies are denoted with the symbol $P_i$ throughout, where the subscript $i$ assumes higher values the greater the distance from Jupiter. Thus, Jupiter itself is denoted with $P_0$, Io with $P_1$, Europa with $P_2$, Ganymede with $P_3$, and Callisto with $P_4$.

Let $\vec{r}_i$ and $\vec{r}_{ij}$ denote the relative vectors $\overline{P_0P_i}$ and $\overline{P_iP_j}$, respectively, with ${r}_i$ and ${r}_{ij}$ indicating the associated Euclidean norms. Furthermore, let $O$ denote the barycenter of the system. By considering the vectorial equality $\vec{r}_i=\overline{OP_i}-\overline{OP_0}$, we can derive the equations describing the dynamics in a Jovicentric frame $(P_0, x, y, z)$. These read

\begin{equation}
\ddot{\vec{r}}_i = \frac{\vec{F}_i}{m_i} - \frac{\vec{F}_0}{m_0}, \quad i=1,\dots,4,
\end{equation}
where $m_i$ and $m_0$ denote the mass of the moon and the mass of Jupiter, respectively, while $\vec{F}_i$ and $\vec{F}_0$ stand for the whole external forces acting on the two masses. With a notation similar to the one adopted for the relative vectors, we indicate with $\vec{F}_{ij}$ the force exerted by $P_j$ over $P_i$. Thus, for example, if we introduce the simplified gravitational potential ${U}_{ij} = {U}_{\bar{i}\bar{j}} = \frac{1}{{r}_{ij}}$, the gravitational attraction exerted by $P_j$ over $P_i$ becomes $\vec{F}_{ij} = Gm_im_j \nabla_i {U}_{ij}$.

In the notation for the gravitational potential we have placed a bar above the indices. This means that the body corresponding to that index is modeled as a point mass. In a similar manner, we model an eventual contribution due to an oblate shape by a hat above the corresponding subscript. For instance, the total force exerted by Jupiter over the $i$th moon can be written as

\begin{equation}
\vec{F}_{i0} = \vec{F}_{\bar{i}\bar{0}} + \vec{F}_{\bar{i}\hat{0}}\ ,
\end{equation}
where $\vec{F}_{\bar{i}\bar{0}}=Gm_im_0 \nabla_i {U}_{\bar{i}\bar{0}}$ and $\vec{F}_{\bar{i}\hat{0}}=Gm_im_0 \nabla_i {U}_{\bar{i}\hat{0}}$. This last term represents the action of the $P_0$ triaxiality over $P_i$, and it can be expressed using equatorial spherical coordinates $({r}_i, \phi_i, \lambda_i)$, where the last two variables denote the latitude and longitude of $P_i$ (the vector length is invariant under rotations), respectively. If we denote by $R$ the equatorial radius of $P_0$, the potential ${U}_{\bar{i}\hat{0}}$ describing the previous action can be written as

\begin{equation}\label{UU}
{U}_{\bar{i}\hat{0}} = {U}_{\bar{i}\hat{0}}^{(1)} + {U}_{\bar{i}\hat{0}}^{(2)}\ ,
\end{equation}
where

\begin{equation}\label{zonalObl}
{U}_{\bar{i}\hat{0}}^{(1)} = \sum_{n=2}^\infty \frac{R^n}{{r}_i^{n+1}} J_n \tilde P_n\left( \sin \phi_i\right)
\end{equation}
and

\begin{equation}\label{tesseralObl}
{U}_{\bar{i}\hat{0}}^{(2)} = \sum_{n=2}^\infty \frac{R^n}{{r}_i^{n+1}} \sum_{m=1}^n \tilde P_n^m\left( \sin \phi_i \right) \left[ C_{nm} \cos m\lambda_i + S_{nm} \sin m\lambda_i \right]\ .
\end{equation}
The quantities $J_n$, $C_{nm}$, and $S_{nm}$ are all constants depending on the particular primary, while $\tilde P_n^m$ denotes the associated Legendre polynomials (with $\tilde P_n$ equal to $\tilde P_n^0$). For an in-depth treatment, see \citet{KaBook} and \citet{CG18}.

In conclusion, and this choice is justified in Section \ref{FigSec}, we can construct a basic dynamical model for the Galilean moons by considering only their mutual gravitational interactions and the influence of Jupiter (modeled as an oblate planet). Consequently, the basic differential system for the $i$th moon becomes

\begin{equation}
\begin{gathered}
\ddot{\vec{r}}_i = -\frac{G(m_0+m_i)}{{r}_i^3}\vec{r}_i + \sum_{j=1,\,j\ne i}^N Gm_j\left( \frac{\vec{r}_j- \vec{r}_i}{{r}_{ij}^3} - \frac{\vec{r}_j}{{r}_j^3} \right)\\
+G(m_0+m_i)\nabla_i U_{\bar{i}\hat{0}} + \sum_{j=1,\,j\ne i}^N Gm_j\nabla_j U_{\bar{j}\hat{0}}\ .
\end{gathered}
\label{motionEqs}
\end{equation}
We remark that unless indicated otherwise, the oblate potentials $U_{\bar{i}\hat{0}}$ are restricted to the zonal terms in Eq. \eqref{zonalObl}, and the corresponding series are truncated at $J_2$. Furthermore, in light of the masses considered, we take the sums up to $N=4$. However, because of the general character of Eq. \eqref{motionEqs}, we prefer to keep an equally general notation, which will facilitate understanding the next subsections. Of course, these equations can be extended by considering other masses or triaxiality contributions (as in \citet{La-A}). Finally, we remark that the last term in Eq. \eqref{motionEqs} and the factor associated with $m_i$ in the preceding equation represent indirect forces resulting from the oblateness of the central body. Numerically, they are relatively small ($O(m_iJ_2)$ at most), but they allow for a better preservation of the total energy of the system. Following \citet{La-A}, in the remaining paper we refer to them as \emph{\textup{additional oblateness forces}}.


\subsection{Precession of Jupiter's line of nodes}\label{PrecSec}

In this subsection we provide the formulas necessary to take into account the precession of Jupiter's mean equatorial plane in the equations above (along with the rotation around the corresponding polar axis). This effect is taken into account solely to represent the evolution of the Laplace angle according to the ephemerides set, and not for the numerical integration of the benchmark model. We still describe it here in order to provide a general form for the oblateness terms introduced in the latter.

The evolution of the Jovian rotation pole and prime meridian line relative to the J2000 frame, without taking into account nutation terms, is constructed from the 2006 IAU report (rotation) and the corresponding 2009 version (as in the Spice toolkit). Specifically, the equations are\footnote{\url{https://naif.jpl.nasa.gov/pub/naif/generic_kernels/pck/pck00010.tpc}}

\begin{equation}\label{PrecFor}
\begin{aligned}
\alpha &= 268.056595 - 0.006499\, T,\\
\delta &= 64.495303 + 0.002413\, T,\\
W &= 284.95 + 870.536\, d.
\end{aligned}
\end{equation}
In the Eq. \eqref{PrecFor}, all written in degrees $\alpha$ stands for the right ascension, $\delta$ for the declination, and $W$ for the longitude of the prime meridian. Furthermore, $T$ and $d$ denote the times in Julian centuries ($36525$ days) and Julian days ($86400$ seconds), respectively, from the standard epoch J2000.

\begin{figure}[t]
\centering
\resizebox{6.5cm}{!}{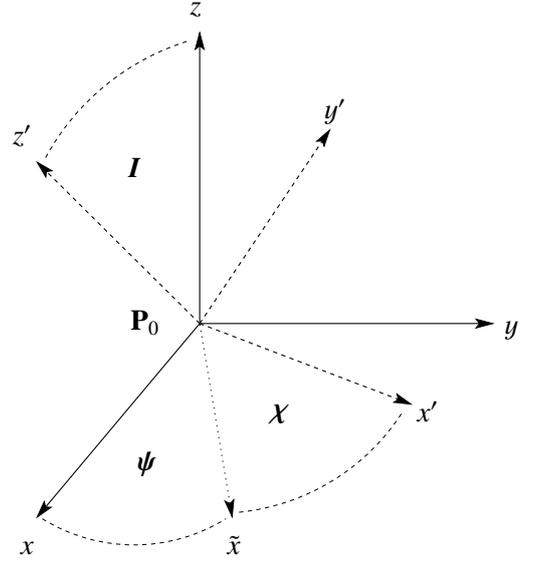}
\caption{Precession and rotation of Jupiter's mean equatorial frame $(P_0,x^\prime,y^\prime,z^\prime)$ with respect to the J2000 frame $(P_0,x,y,z)$, represented through the Euler angles $(\psi, I, \chi)$.}
\label{Prc_Frame}
\end{figure}

Through the angles above, we can define the rotation from the J2000 frame to the one individualized by the directions, at a given epoch of Jupiter's mean north pole and equatorial node. This is done, as shown in Figure \ref{Prc_Frame}, by considering the Euler angles

\begin{equation}
I = 90 - \delta,\ \ \ \psi = \alpha + 90,\ \ \ \chi = W.
\label{eulAng}
\end{equation}
In turn, these angles define the intrinsic rotation  $z-\tilde x-\tilde z^\prime$, with $\tilde z^\prime$ coinciding with $z^\prime$ in the figure. This rotation is defined by the matrix (see for example \citet{LaThesis})

\begin{multline}
M = \left(
\begin{matrix}
\cos\chi \cos\psi - \sin\chi \sin\psi \cos I  \\
-\sin\chi \cos\psi - \cos\chi \sin\psi \cos I  \\
\sin\psi \sin I  \\
\end{matrix}\right.
\\
\left.
\begin{matrix}
\cos\chi \sin\psi + \sin\chi \cos I \cos\psi  &  \sin\chi \sin I\\
-\sin\chi \sin\psi + \cos\chi \cos I \cos\psi  &  \cos\chi \sin I\\
-\sin I \cos\psi  &  \cos I
\end{matrix}\right)
.
\end{multline}

\subsection{Gradient of the oblate potential}\label{OblSec}

In order to implement Eq. \eqref{motionEqs}, we need an explicit, general expression for the term $U_{\bar{i}\hat{0}}$, which has to be valid also when the precession of Jupiter's mean equatorial plane is taken into account. The formulas that we provide here can be found in \cite{LaThesis}, along with a more extensive discussion of their derivation.

\begin{table*} 
\centering
\begin{tabular}{cccc}
\hline
\hline

Galilean moon   &$x\ (\dot x)$ &$y\ (\dot y)$ &$z\ (\dot z)$\\

\hline

 Positions (Io)   &3.9560614914644e+05  &1.4174195472057e+05    &1.8927919180829e+02\\
 Velocities (Io)  &-5.9016124704240e+00 &1.6368931345219e+01    &-8.3134391183429e-03\\
 Positions (Europa) &-5.5008678962774e+05       &-3.7344141766412e+05   &3.0053327179017e+03\\
 Velocities (Europa) &7.8187490244338e+00       &-1.1454811869792e+01   &9.2524448980401e-02\\
 Positions (Ganymede) &-8.0001364982337e+05     &-7.1100915843111e+05   &2.0066504100776e+03\\
 Velocities (Ganymede) &7.2399249841294e+00     &-8.1226251713938e+00   &-3.3603495285878e-02\\
 Positions (Callisto) &2.6813508156045e+05      &1.8624715507078e+06    &-6.3762880769717e+03\\
 Velocities (Callisto) &-8.1151884687706e+00    &1.2298492010811e+00    &6.4560951005827e-03\\

\hline

\end{tabular}
\vspace*{1mm}
\caption{SPICE Cartesian elements for the Galilean moons in the Jovicentric equatorial frame. The date is J2000. Positions are in kilometers, and velocities are in kilometers per second.}
\label{table:1}
\end{table*}
Limiting ourselves to zonal harmonics (which are the fundamental harmonics here), we first observe that in light of the previous section, we have

\begin{equation}\label{SinLat}
\sin\phi_i = \frac{{z}_i^\prime}{{r}_i} =\frac{{x}_i \sin I \sin\psi - {y}_i \cos\psi \sin I + {z}_i \cos I}{{r}_i}.
\end{equation}
If we denote with $\gamma_i$ alternatively each of the variables $(x_i, y_i, z_i)$, then we have

\begin{equation}
\frac{\partial \sin\phi_i}{\partial \gamma_i} = \frac{1}{{r}_i}\left( \frac{\partial {z}_i^\prime}{\partial \gamma_i} - \frac{\gamma_i \sin\phi_i}{{r}_i} \right)
\end{equation}
and also

\begin{equation}
\begin{split}
\frac{\partial}{\partial \gamma_i} \left[ \frac{\tilde P_n(\sin\phi_i)}{{r}_i^{n+1}} \right] = \frac{1}{{r}_i^{n+1}} \Bigg[ \frac{\partial \sin\phi_i}{\partial \gamma_i} &\frac{d\tilde P_n(\sin\phi_i)}{d\sin\phi_i} \\
&- \frac{\tilde P_n(\sin\phi_i)(n+1)\gamma_i}{{r}_i^2} \Bigg].
\end{split}
\end{equation}
By joining the previous equalities, we obtain for the derivatives of Eq. \eqref{zonalObl} the expression

\begin{equation}
\begin{split}
\frac{\partial {U}_{\bar{i}\hat{0}}^{(1)}}{\partial \gamma_i} = - \sum_{n=2}^\infty \frac{R^nJ_n}{{r}_i^{n+2}} \Bigg[ \Bigg( \frac{\partial {z}_i^\prime}{\partial \gamma_i} - &\frac{\gamma_i \sin \phi_i}{{r}_i} \Bigg) \frac{d\tilde P_n(\sin \phi_i)}{d\sin \phi_i} \\
&- \frac{(n+1) \gamma_i \tilde P_n(\sin \phi_i)}{{r}_i}\Bigg].
\end{split}
\end{equation}
In conclusion, by limiting ourselves to consider only the harmonic $J_2$, we obtain explicitly

\begin{equation}
\begin{aligned}
\frac{\partial {U}_{\bar{i}\hat{0}}^{(1)}}{\partial {x}_i} &= \frac{R^2 J_2}{{r}_i^4}\left[ \frac{{x}_i}{{r}_i}\left( \frac{15}{2}\sin^2\phi_i - \frac{3}{2} \right) - 3\sin\phi_i \sin I \sin\psi \right],\\
\frac{\partial {U}_{\bar{i}\hat{0}}^{(1)}}{\partial {y}_i} &= \frac{R^2 J_2}{{r}_i^4}\left[ \frac{{y}_i}{{r}_i}\left( \frac{15}{2}\sin^2\phi_i - \frac{3}{2} \right) + 3\sin\phi_i \sin I \cos\psi \right],\\
\frac{\partial {U}_{\bar{i}\hat{0}}^{(1)}}{\partial {z}_i} &= \frac{R^2 J_2}{{r}_i^4}\left[ \frac{{z}_i}{{r}_i}\left( \frac{15}{2}\sin^2\phi_i - \frac{3}{2} \right) - 3\sin\phi_i \cos\psi \right],
\end{aligned}
\end{equation}
which can be further expanded by plugging them into Eq. \eqref{SinLat}.


\subsection{Conservation of the energy}\label{NumConsSec}

The equations of motion, Eq. \eqref{motionEqs}, are propagated through the use of an adaptive Runge-Kutta algorithm of seventh order, with initial conditions given by the ephemerides of the moons at the date J2000. Since in Spice these are available only up to about 2100, we propagate the conditions only up to 100 years (so that we can also compare our results to those of \citet{La-A}). 

\begin{figure}[t]
\centering
\includegraphics[width=.75\columnwidth]{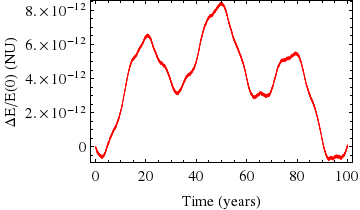}
\caption{Energy relative error (in normalized units) committed when propagating Eq. \eqref{motionEqs}. Units are defined by Io's mean orbital period and semi-major axis, and by Jupiter's mass.}
\label{enrgErr}
\end{figure}

To verify the correctness of the numerical integration, we considered the temporal evolution of the total energy of the system. If we denote by $M$ the sum of the system masses, then we can express the energy $E$ as \citep{PR17}

\begin{equation}
E = \sum_{i=1}^N \frac{m_i\dot{{\bm r}}_i^2}{2} - \frac{1}{2M}\left( \sum_{i=1}^N m_i\dot{{\bm r}}_i \right)^2 - U\ ,
\end{equation}
where $U$ is the potential of the system. In light of the contributions considered, we can write this term as

\begin{equation}
U = \sum_{i=1}^N Gm_im_0\left( \frac{1}{{r}_i} + {U}_{\bar{i}\hat{0}} \right) + \sum_{i=1}^{N-1}\sum_{j=i+1}^N \frac{Gm_jm_i}{{r}_{ij}}\ ,
\end{equation}
where the terms ${U}_{\bar{i}\hat{0}}$ are defined in Eq. \eqref{UU} and again $N=4$. Figure \ref{enrgErr} shows that the energy remains well preserved for the entire timescale we considered.


\subsection{Results}\label{FigSec}

We evaluate here the effectiveness of the Cartesian model we introduced in capturing the librating expressions associated with the Laplace resonance. We remark that since our interest is mainly analytical, our accuracy thresholds are fairly relaxed. Furthermore, in Table \ref{table:1} we provide the values of the initial conditions used to generate the plots of this subsection.

\begin{figure}[t]
\centering
\includegraphics[width=.75\columnwidth]{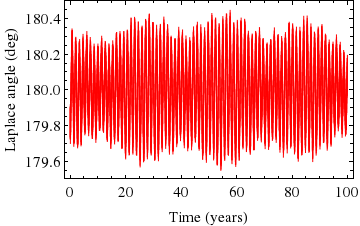}\\[.15cm]
\includegraphics[width=.75\columnwidth]{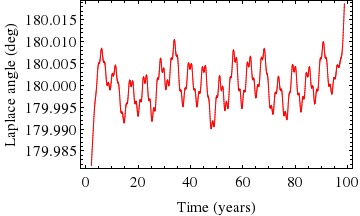}
\caption{{\bf Top:} Laplace argument ephemerides history from the J2000 epoch, including the effect of Jupiter's mean equatorial plane precession. {\bf Bottom:} Result after a low-pass filter at $1000$ days (tails are a byproduct of the procedure). Ephemerides from the NASA Spice toolkit.}
\label{ephLapl}
\end{figure}

\begin{table} 
\label{table:3}
\centering
\begin{tabular}{cc}
\hline
\hline

Parameter       & Value\\

\hline

 Mass (Jupiter)   &1.8986e+27\\
 Mass (Io)   &893.3e+20\\
 Mass (Europa)  &479.7e+20\\
 Mass (Ganymede) &1482.0e+20\\
 Mass (Callisto) &1076.0e+20\\
 $J_2$ &1.478e-2\\
 G   &6.67259e-20\\

\hline

\end{tabular}
\vspace*{1mm}
\caption{Parameters for the simulations: masses are in kilograms, and the gravitational constant has kilometers as the unit of length (the others are SI units).}
\end{table}
As a first step, we plot in the top panel of Figure \ref{ephLapl} the history of the Laplace argument as derived from the ephemerides. We do not show it here, but taking the power spectrum of this signal reveals several frequencies with higher intensity than the frequency we are interested in (which occurs slightly later than $2000$ days). A low-pass filter at $1000$ days is enough to remove these higher frequencies, leaving a signal with a maximum amplitude of about $0.02$ degrees. Not only is this a lower value than reported by \citet{Li-B} and \citet{MuVaMoSc}, but when we compare our plot with Figure $2$ in \citet{MuVaMoSc}, we still have higher frequencies (again, this can be confirmed by taking a fast Fourier transform). These discrepancies may be due to several factors, ranging from the different set of ephemerides to the function used for the filtering. For reference, our procedure is performed automatically in Mathematica 11, using the default options.

The top panel of Figure \ref{intLapl} shows that the libration of the Laplace angle, complete with the desired period, appears already when we restrict our Cartesian model to only purely gravitational terms. The amplitude is extremely large, however: about $70\degree$. As previously argued, to recover the (almost) correct history of the angle, it is sufficient to plug in the model Jupiter's oblateness potential truncated at the zonal harmonic $J_2$. The second (integration) and third (ephemerides) panels of Figure \ref{intLapl} show that the approximation is quite good for the timescale we considered ($100$ years). We remark, however, that the variation increases from $0.05\degree$ to $0.15\degree$ degrees with a modulation coherent with the libration main frequency, as is apparent in the bottom plot of the same figure. This suggests that additional effects acting on the mean longitudes may have to be taken into account for longer time spans.

As an additional check, useful mainly for the next sections, we estimate the linear combinations of observed mean motions corresponding to the quasi-resonant interactions Io-Europa and Europa-Ganymede. These can be computed either via a fast Fourier transform or by numerical estimation of the periods. The first method is heavily influenced by the sampling frequency, particularly for the fast-traveling Io, thus here we rely on the second method.

We determined for each moon the orbital period $T$ by considering the first return of the mean longitude to its initial value, and derived the observed mean motion as $2\pi/T$. This computation was then repeated over $500$ periods and the average was taken as reference value. The diagnostics corresponding to these values (for model \eqref{motionEqs}) are

\begin{equation}
\begin{aligned}
n_1 - 2n_2 &=0.739926\ \degree / day,\\
n_2 - 2n_3 &=0.740128\ \degree / day,
\end{aligned}
\end{equation}
and the error on the nominal values \eqref{laplMM} is $O(10^{-4})$ degrees per day.

%
%

\section{Hamiltonian formulation}


\subsection{Introduction}

\begin{figure}[t]
\centering
\includegraphics[width=.75\columnwidth]{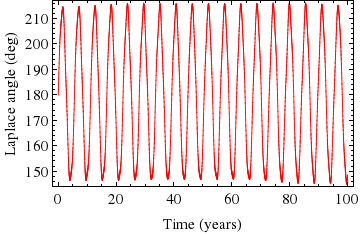}\\[.15cm]
\includegraphics[width=.75\columnwidth]{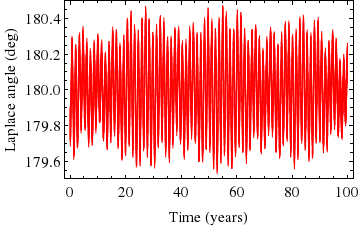}\\[.15cm]
\includegraphics[width=.75\columnwidth]{eph_lapl_eqt.png}\\[.15cm]
\includegraphics[width=.75\columnwidth]{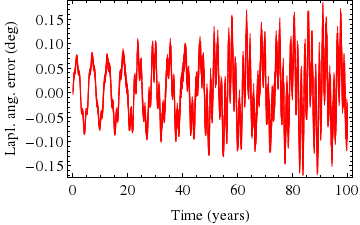}
\caption{{\bf Top to bottom:} Integration history of the Laplace argument obtained without oblateness terms, including Jupiter's $J_2$ harmonic, the ephemerides history of the argument in the same time span, and the difference of the last two plots. The starting epoch is J2000, with initial conditions given by the corresponding ephemerides.}
\label{intLapl}
\end{figure}
In this section we introduce a Hamiltonian model for the problem at hand. The utility of doing so is well understood, and derives from several tools developed for this formulation that allow for a deeper understanding of the underlying dynamics. A complete discussion on this is beyond the scope of the paper and is reserved for parallel works (see \citet{CePaPu} and \citet{CePaPu-B}). Here we focus on presenting a basic version of this model and on validating it by comparing its results with the benchmark Cartesian model.

The section is organized as follows. In Section \ref{jovAndJac} we lay down our Hamiltonian in osculating orbital elements and introduce series expansions in the neighborhoods of the two near 2:1 mean motion resonances comprising the Laplace resonance, and retain only a finite number of resonant terms. The derivation is then completed in Section \ref{theRest}, where we introduce expressions for the perturbative effects and the canonical set of coordinates defining the shape of the function. Finally, in Section \ref{resCoord} we transform this Hamiltonian by using an appropriate set of resonant coordinates to focus our attention on the resonant variables of the system.


\subsection{Hamiltonian in orbital elements}\label{jovAndJac}

As stated before, we do not discuss the derivation of the Hamiltonian, since this is done in \citet{CePaPu}, and more details are provided in \citet{CePaPu-B}. Instead, we look at its components and provide the series expansions in orbital elements leading to its final form.

With the exclusion of oblateness effects, which we examine in the next subsection, we include in our Hamiltonian only the mutual gravitational contributions of Jupiter, Io, Europa, and Ganymede. As we show below, the choice of excluding the full influence of Callisto is a necessary step to facilitate some of the calculations, although it has consequences on some aspects of the dynamics.

In order to describe the Hamiltonian, we introduce the auxiliary variables

\begin{equation}
\begin{gathered}
M_1=m_0+m_1,\qquad \mu_1={{m_0m_1}\over M_1},\qquad \kappa_1={m_1\over M_1},\\
M_2=M_1+m_2,\qquad \mu_2={{M_1m_2}\over M_2},\qquad \kappa_2={m_2\over M_2},\\
M_3=M_2+m_3,\qquad \mu_3={{M_2m_3}\over M_3},\qquad \kappa_3={m_3\over M_3}.\\
\end{gathered}
\end{equation}
As described in \citet{Ma}, a possible way to introduce a Hamiltonian for the system is to use Jacobi coordinates. In addition to the motion of the barycenter, this leads  to three main contributions: an unperturbed part generated by Jupiter's pull, and a perturbation due to the mutual interactions of the satellites, which we can divide into a direct and an indirect part (see \citet{MuDe}, for example). As shown in \citet{CePaPu-B}, these include terms in the variables $\kappa_i$, which can be expanded to first order into such variables to obtain a Jovicentric approximation of the Hamiltonian.

The latter can be written in terms of the osculating orbital elements as follows. If we discard the motion of the barycenter (which is constant), the unperturbed part can be separated into three unperturbed two-body energies as

\begin{equation}
H_K=-{{GM_1\mu_1}\over {2a_1}}-{{GM_2\mu_2}\over {2a_2}}-{{GM_3\mu_3}\over {2a_3}},
\label{kepHam}
\end{equation}
where $a_i$ denotes the osculating semi-major axis of the $i$th moon. For the direct and indirect terms of the perturbation, we can work in a neighborhood of the exact 2:1 mean motion commensurabilities corresponding to the quasi-resonant interactions Io-Europa and Europa-Ganymede and, after some transformations, obtain an expansion up to second order in the eccentricities. We remark that in this expansion we chose to retain only low-frequency terms associated with the resonant angles. Furthermore, it is important to note that we are working with two separate quasi-resonances constituting the Laplace resonance, and not with the latter itself. As we show in Section \ref{resCoord}, it is possible to introduce this argument with a suitable canonical change of variables.

We obtain for the perturbing terms the formulae (as mentioned, compare with \citet{Ma} and \citet{MuDe})

\begin{equation}\label{expHam}
\begin{aligned}
&H_P^{1,2} = -{{Gm_1m_2}\over a_2}\big(B_0(\alpha_{1,2})+f_1^{1,2}e_1\cos(2\lambda_2-\lambda_1-\varpi_1) \\
&+f_2^{1,2}e_2\cos(2\lambda_2-\lambda_1-\varpi_2) + f_3^{1,2}(e_1^2+e_2^2) \\
&+ f_4^{1,2} e_1 e_2\cos(\varpi_1 - \varpi_2) + f_5^{1,2} e_1 e_2 \cos( 4\lambda_2 -2\lambda_1 -\varpi_1 -\varpi_2) \\
&+ f_6^{1,2} e_1^2 \cos(4\lambda_2 -2\lambda_1 -2\varpi_1) + f_7^{1,2} e_2^2 \cos(4\lambda_2 - 2\lambda_1 -2\varpi_2)\big), \\[.1cm]
%
&H_P^{2,3} = -{{Gm_2m_3}\over a_3}\big(B_0(\alpha_{2,3})+f_1^{2,3}e_2\cos(2\lambda_3-\lambda_2-\varpi_2) \\
&+f_2^{2,3}e_3\cos(2\lambda_3-\lambda_2-\varpi_3) + f_3^{2,3}(e_2^2+e_3^2) \\
&+ f_4^{2,3} e_2 e_3\cos(\varpi_2 - \varpi_3) + f_5^{2,3} e_2 e_3 \cos( 4\lambda_3 -2\lambda_2 -\varpi_2 -\varpi_3) \\
&+ f_6^{2,3} e_2^2 \cos(4\lambda_3 -2\lambda_2 -2\varpi_2) + f_7^{2,3} e_3^2 \cos(4\lambda_3 - 2\lambda_2 -2\varpi_3)\big), \\[.1cm]
%
&H_P^{1,3} = -{{Gm_1m_3}\over a_3}\big(B_0(\alpha_{1,3}) + f_3^{1,3}(e_1^2+e_3^2) \\
&+ f_4^{1,3} e_1 e_3\cos(\varpi_1 - \varpi_3) \big).
\end{aligned}
\end{equation}
The functions $f_k^{i,j}$ are linear combinations of the \emph{\textup{Laplace coefficients}} $b_s^{(n)}$ and the associated derivatives (see, e.g., \citet{MuDe} or \citet{ShMa}), which in turn are functions of the semi-major axis ratios $\alpha_{i,j}={a_i\over a_j}$. The terms $B_0(\alpha_{i,j})$ equal ${1\over 2}b_{1/2}^{(0)}(\alpha_{i,j})-1$. We also note that all the inclinations have been set to zero, and therefore the model we consider is planar. This assumption is reasonable (although it has some consequences) since the inclinations are on the order of $10^{-1}$ degrees at most.

\begin{table*} 
\begin{tabular}{ccccc}
\hline
\hline

Element &Io &Europa &Ganymede &Callisto\\

\hline
 
 Semi-major axis &4.2203882962178e+05 &6.7125250226694e+05 &1.0705037570640e+06 &1.8827839858043e+06\\
 Eccentricity &4.7208180352021e-03 &9.8185357623597e-03 &1.4579320486125e-03 &7.4398609033480e-03\\
 Inclination &3.7583968915495e-02 &4.6224994231893e-01 &2.0686778845732e-01 &1.9964604311274e-01\\
 Longitude of the node &2.4307740400984e+02 &1.8009776470137e+02 &7.2912129534872e+01 &1.5833629680670e+02\\
 Argument of the perijove &1.6168078566908e+02 &4.8941393235721e+01 &2.3160580130411e+02 &1.9754711726587e+02\\
 Mean anomaly &3.3518221112606e+02 &3.4542002854049e+02 &2.7727663735805e+02 &8.5074032871365e+01\\

\hline
\end{tabular}
\vspace*{1mm}
\caption{Osculating orbital elements corresponding to the Cartesian conditions in Table \ref{table:1}. Semi-major axes are in kilometers, and the angles are expressed in degrees. The epoch is J2000.}
\label{table:2}
\end{table*}


\subsection{Canonical elements and the perturbative effects}\label{theRest}

To conclude our construction, we need to select an appropriate set of canonical variables in which to express the Hamiltonian above. For this work we consider the modified Delaunay elements defined by the equations \citep{Ma}

\begin{equation}\label{delEm}
\begin{gathered}
\lambda_i, \quad L_i = \mu_i \sqrt{G M_i a_i}\simeq m_i \sqrt{G m_0 a_i},\\
-\varpi_i, \quad G_i\equiv L_i (1-\sqrt{1-e_i^2})\simeq \frac{1}{2}L_i e_i^2.
\end{gathered}
\end{equation}
Because we consider only three moons, the final Hamiltonian has six degrees of freedom. Thanks to our choice in the terms to retain, and with the change of coordinates in the next section, it is possible to reduce the complexity of this Hamiltonian even more.

The same type of steps that we have taken for the gravitational part of the Hamiltonian has to be considered for the perturbations taken into account. For example, we showed in the previous section that Jupiter's oblateness potential, truncated at the $J_2$ harmonic, is the single most important factor for recovering the amplitude of the Laplace argument. Since we are interested in recovering only the latter's main frequency, for the Hamiltonian model we only consider the secular part of the perturbation provided by the $J_2$ harmonic. This can be expressed in terms of the osculating orbital elements as

\begin{equation}\label{j2Ham}
H_{J}=-\sum_{i=1}^3{{GM_i\mu_i}\over {2a_i}}\ \left[J_2({R\over a_i})^2(1+{3\over 2}e_i^2)-{3\over 4}\right],
\end{equation}
and is trivially converted via the canonical elements introduced before. Similarly, we can reintroduce Callisto as a perturbation. If we only consider its secular part, we have

\begin{equation}\label{CsHam}
H_{C}=-\sum_{i=1}^3{{Gm_im_4}\over {a_4}}\ \left[B_0(\alpha_{i,4}) + {\alpha_{i,4}\over 8}b_{3/2}^{(1)}(\alpha_{i,4})(e_i^2+e_C^2)\right].
\end{equation}
Unless explicitly mentioned, this last perturbation is not considered in the Hamiltonian model, since the Laplace resonance does not directly involve Callisto.


\subsection{Resonant coordinates}\label{resCoord}

As described above, we can introduce a canonical change of coordinates to highlight the Laplace argument as a variable of the system. This is defined in terms of the modified Delaunay elements \eqref{delEm} as

\begin{equation}\label{resVar}
\begin{aligned}
q_1 &= 2\lambda_2-\lambda_1-\varpi_1, \quad\quad P_1 = G_1,\\
q_2 &= 2\lambda_2-\lambda_1-\varpi_2, \quad\quad P_2 = G_2,\\
q_3 &= 2\lambda_3-\lambda_2-\varpi_3, \quad\quad P_3 = G_3,\\
q_4 &= 3\lambda_2-2\lambda_3-\lambda_1, \quad \;\;\; P_4 = {\scriptstyle{\frac13}}\left(L_2-2(G_1+G_2)+G_3\right),\\
q_5 &= \lambda_1-\lambda_3, \quad\quad\quad\quad \;\;\; P_5 = {\scriptstyle{\frac13}}\left(3L_1+L_2+G_1+G_2+G_3\right),\\
q_6 &= \lambda_3, \quad\quad\quad\quad\quad\quad \;\;\; P_6 = L_1+L_2+L_3-G_1-G_2-G_3.
\end{aligned}
\end{equation}
Equation \eqref{expHam} shows that with the terms that we decided to retain in our expansions, the variables $q_5$ and $q_6$ do not appear in the Hamiltonian ($P_6$ is the total angular momentum \citep{LaPe} of the system). Under the previous change of coordinates, the Hamiltonian $H(q_i, P_i)$ therefore does possess four d.o.f.

Furthermore, and this is explained in greater detail in \citet{CePaPu} or \citet{BrHa}, it can be proved that the Hamiltonian, with the gravitational terms restricted to first order in the eccentricity, has exactly $\text{eight}$ equilibrium configurations (excluding the symmetries of the system). The only stable configuration, whose discovery is generally attributed to de Sitter \citep{Si}, is given by

\begin{equation}\label{stbEq}
\left\{ q_1, q_2, q_3, q_4\right\} = \left\{ 0, \pi, \pi, \pi \right\}.
\end{equation}
This equilibrium does not correspond to the actual state of the system, since we know from observations that the angle $q_3$  rotates. This is an important detail and is exploited in the next section to facilitate the search for the correct amplitude of the resonant argument.

%
%

\section{Numerical simulations}\label{CmpSim}


\subsection{Comparison of the models}

\begin{figure}[t]
\centering
\includegraphics[width=.75\columnwidth]{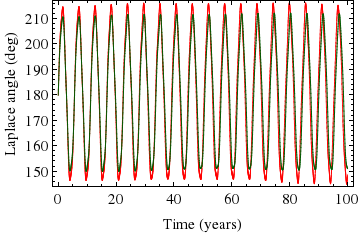}\\
\includegraphics[width=.75\columnwidth]{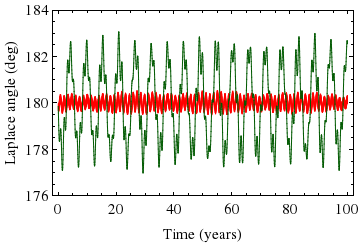}
\caption{{\bf Top:} Laplace argument integration history obtained without oblateness terms. {\bf Bottom:} Same evolution with the addition of the $J_2$ terms. In red we show the results from integration of the\textcolor[rgb]{0,0,0}{ \textcolor[rgb]{0,0,0}{{\color{red}} }}Cartesian model, in green equivalent from the Hamiltonian.}
\label{hamLapl}
\end{figure}
Here we numerically compare the Cartesian and Hamiltonian formulations in order to measure how well the latter approximates the main amplitude and frequency of the Laplace argument, and the Jovian moons' dynamics (except for Callisto, of course). The initial conditions are the same for the two models and correspond to the ephemerides at the date J2000, which we also used in Section \ref{FigSec}. For the Hamiltonian model, they are expressed in terms of the osculating orbital elements, whose values are listed in Table \ref{table:2}.

As a first step, in Figure \ref{hamLapl} we superimpose the Hamiltonian evolution of the Laplace argument on the Cartesian evolution presented in Figure \ref{intLapl}. In the top panel, where we exclude the oblateness contribution for both models, the difference is relatively small, both in period and amplitude. However, if we retain the $J_2$ terms (bottom panel), the
corresponding amplitude is out of scale of at least an order of magnitude, even though the main period is still well determined by the Hamiltonian formulation. Of course, it is to be expected that the same initial conditions do not lead to the same evolution of the argument, and it is still possible to recover the correct amplitude, if with some compromise. We show this in the next subsection.

\begin{figure}[t]
\centering
\includegraphics[width=.75\columnwidth]{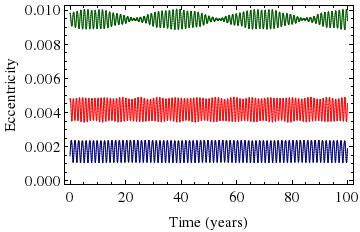}\\[.15cm]
\includegraphics[width=.75\columnwidth]{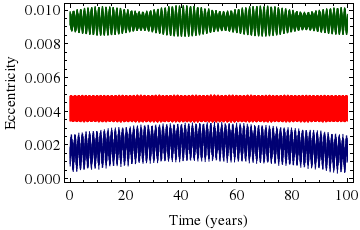}
\caption{{\bf Top:} Jovian moons' eccentricity evolution as computed with the Hamiltonian model. {\bf Bottom:} Corresponding histories derived from the Cartesian model The moons are color-coded (Io in red,  Europa in green, and Ganymede in blue).}
\label{hamEcc}
\end{figure}
We examine the Jovian moons' dynamics stemming from the Hamiltonian's approximation in more detail. In particular, we consider the ``shape'' of the orbit, that is, the evolution of the moons' eccentricities and semi-major axes. In Figure \ref{hamEcc} we plot the history of the former for the Hamiltonian formulation (top panel, $H_C$ not included) and the Cartesian one (bottom panel). In all cases, the absolute error achieves a maximum value of $O(10^{-3})$. In turn, this implies different values for the maximum relative error, with peaks of $40\%$ for Io, a steady increase of up to $15\%$ for Europa, and occasional ventures beyond $100\%$ for Ganymede (although the average is about $50-60\%$). Furthermore, it is evident that an additional frequency in Ganymede's eccentricity evolution is present in the Cartesian case. Particularly for this last effect, the primary suspect might be Callisto, whose gravitational contribution is not taken into account in Figure \ref{hamEcc}. Introducing its secular effect $H_C$ in the Hamiltonian formulation does not significantly change the results, however. Thus, it seems that this lack of accuracy is a price to pay if the full effect of Callisto is not taken into account (and therefore the number of d.o.f. is kept small).

\begin{figure}[t]
\centering
\includegraphics[width=.75\columnwidth]{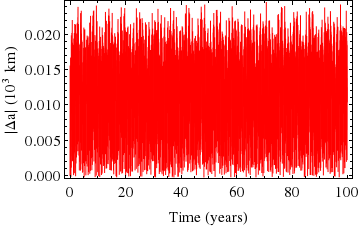}\\[.15cm]
\includegraphics[width=.75\columnwidth]{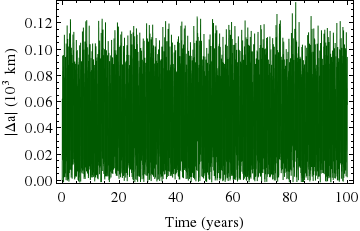}\\[.15cm]
\includegraphics[width=.75\columnwidth]{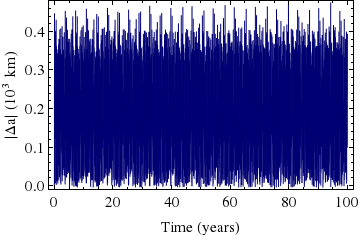}
\caption{History of the absolute errors on the Jovian moons' semi-major axes as computed in the Cartesian and Hamiltonian formulations. The moons are color-coded  (Io in red,  Europa in green,
and Ganymede in blue).}
\label{hamSm}
\end{figure}
In Figure \ref{hamSm} we consider the absolute error of the history of the semi-major axes as obtained from the two formulations. By dividing them for the semi-major axis values obtained with the Cartesian integration, it is easy to check that the relative error committed has an upper bound of about $0.001$ \%. Naturally, since we approximate the real system with a planar system and the inclinations are more or less on the same order, the largest absolute error is committed for Ganymede, which is the moon most distant from Jupiter.

\begin{figure}[t]
\centering
\includegraphics[width=.75\columnwidth]{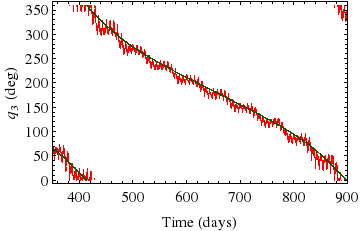}\\[.15cm]
\includegraphics[width=.75\columnwidth]{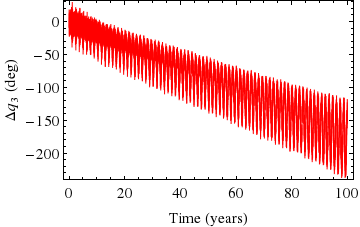}\\[.15cm]
\includegraphics[width=.75\columnwidth]{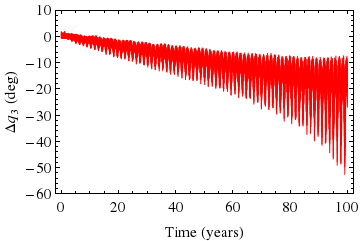}
\caption{{\bf Top to bottom:} {\color{red}} Cartesian and Hamiltonian{\color{darkgreen}} histories for the angle $q_3$, corresponding error on the evolution, and similar error between the Cartesian formulation and the ephemerides set.}
\label{q3Diff}
\end{figure}

Finally, in Figure \ref{q3Diff} we analyze the different model histories for the angle $q_3$ , whose character, as stated in Section 3.4 and as shown in the next subsection, determines whether we are in the Laplace resonance. In the first case, this angle rotates with a period of about $485$ days, as apparent in the top plot from its Cartesian evolution (red). Small period oscillations are present, but they are averaged out when we consider the same angle in the Hamiltonian case (green). Furthermore, as shown in the middle plot, the agreement of the two models on this angle deteriorates with time, which implies a small error in the related drift rate for the two formulations. We remark that as the bottom plot suggests, a similar, less severe displacement exists between the Cartesian model and the ephemerides set. Moreover, these accuracy losses are comparable in terms of magnitude of the relative error with those exhibited by the Laplace argument, as it is apparent by matching the bottom panels of Figure \ref{intLapl} and Figure \ref{hamLapl} with the equivalent panels in Figure \ref{q3Diff}.  

For reference, points in the plots are taken each day, with the exception of the pictures for the semi-major axes, where they are taken every $10$ days, and the top plot of Figure \ref{q3Diff}, where they are taken every hour. This convention is also maintained in the next subsection, where we show why the main resonant behavior is not recovered and how this can be remedied.


\subsection{Conditions for a ``correct'' Hamiltonian resonance}

We showed in the previous section that taking the ephemerides as initial conditions for the Hamiltonian model does not lead to a correct approximation of the Laplace argument amplitude. Similarly, the observed quasi-resonant mean motion combinations slightly disagree with the nominal values \eqref{laplMM}. In particular, the average over a period of 100 years gives

\begin{equation}\label{dgnOrHam}
\begin{aligned}
n_1 - 2n_2 &= 0.734517\ \degree / day,\\
n_2 - 2n_3 &= 0.734591\ \degree / day,
\end{aligned}
\end{equation}
with an error of about $0.005\ \degree / day$. This is reasonable, since there are several approximations in the Hamiltonian model, ranging from its planarity to the more important restrictions on the selected resonant terms.

Fortunately, we can exploit its dynamics, as seen from the lens of the resonant variables in Eq. \eqref{resVar}, to revert the problem and determine the initial conditions that generate values closer to the nominal ones. The first step, as shown in \citet{CePaPu}, is to restrict our resonant Hamiltonian to first order in the eccentricity in the gravitational terms, and retain the full secular contribution of Jupiter's oblateness in Eq. \eqref{j2Ham}, so that we can employ normal form techniques to approximate the dynamics on the phase plane $(q_3,P_3)$ (similarly to \citet{He} and \citet{BCCP}).

\begin{figure}[t]
\centering
\includegraphics[width=.75\columnwidth]{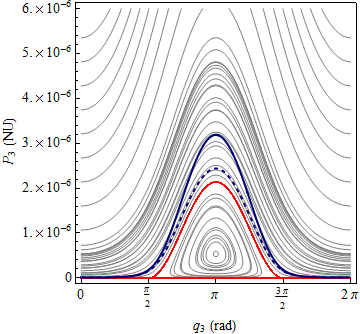}
\caption{Phase plane for the normal form Hamiltonian obtained by restricting Eq. \eqref{expHam} to first-order terms in the eccentricity and retaining the full secular effect of Jupiter's $J_2$ harmonic in Eq. \eqref{j2Ham} (along with Eq. \eqref{kepHam}, of course). The full blue curve is obtained by integrating Hamilton's equations with Eq. \eqref{deSoscel} as initial conditions, except for $q_3 = 0$ and $P_3$ as in Eq. \eqref{lapP3}. These are used also for the dashed blue curve, where Eq. \eqref{CsHam} and the second-order terms of Eq. \eqref{expHam} are also taken into account.}
\label{deSplane}
\end{figure}

As mentioned in Section \ref{resCoord}, the first-order Hamiltonian has exactly one stable equilibrium, and the difference between this (``de Sitter'') equilibrium and the actual state of the system lies in the rotating character of the angle $q_3$. Correspondingly, in Figure \ref{deSplane}, where the phase space of the normal form Hamiltonian is plotted, the first appears as an elliptic fixed point, while the second lies in the vicinity of one of the rotational tori (blue curves). The two blue curves are both computed from direct integration of Hamilton's equations; the full curve corresponds to the sum of the terms in Eqs. \eqref{kepHam}, \eqref{expHam} (restricted to first order in the eccentricity), and \eqref{j2Ham}, all translated into the coordinates of Eq. \eqref{delEm}. The dashed curve instead adds to the previous addenda the terms at second order in the eccentricity for Eq. \eqref{expHam} and the secular contribution in Eq. \eqref{CsHam} of Callisto. The red curve is outlined only to remark its boundary status between the rotational and librational regimes (it is not a formal separatrix). The normal form Hamiltonian associated with the phase plane is given by

\begin{equation}
\begin{aligned}
&H_{dS} = -0.003638 P_3 - 1.6497 P_3^2\\
&- 6.1362\times 10^{-6} \sqrt{P_3} \cos{q_3} + 1.1155\times 10^{-5} P_3^{3\over 2} \cos{q_3},
\end{aligned}
\end{equation}
where the units of measure are determined by Io's semi-major axis, its mean orbital period in days, and by Jupiter's mass.

The blue curves were determined by starting from the numerical location of the de Sitter equilibrium, which is obtained from Hamilton's equations evaluated at $\left\{ q_1, q_2, q_3, q_4\right\} = \left\{ 0, \pi, \pi, \pi \right\}$ by employing a root-finding algorithm for the momenta $P_i$. Differently from \citet{CePaPu}, for this computation we derived the de Sitter equilibrium by also including the terms at second order in the eccentricity in the gravitational terms (those pertaining to the oblateness remained unchanged). The values obtained in this way were subsequently converted into Delaunay elements $L_i$ and $G_i$. From these, we can derive the approximate location of the equilibrium in terms of the orbital elements, that is,

\begin{equation}
\begin{aligned}
a_1^\star &= 422043.201 \; {\rm km}, \quad e_1^\star = 0.00422405,\\
a_2^\star &= 671235.280 \; {\rm km}, \quad e_2^\star = 0.00948442,\\
a_3^\star &= 1070501.267 \; {\rm km}, \quad e_3^\star = 0.00063468.
\end{aligned}
\label{deSoscel}
\end{equation}
Furthermore, we point out that in addition to having been computed with additional terms in the Hamiltonian, these values differ from those in \citet{CePaPu} because the initial conditions there are taken from \citet{La-A}.

\begin{figure}[t]
\centering
\includegraphics[width=.75\columnwidth]{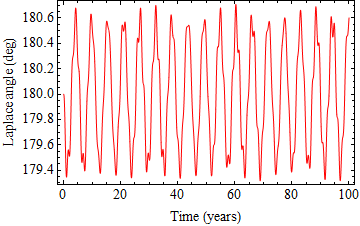}\\[.15cm]
\includegraphics[width=.75\columnwidth]{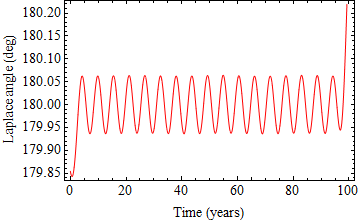}
\caption{{\bf Top:} Hamiltonian history of the Laplace argument obtained from the initial conditions of the de Sitter equilibrium, Eq. \eqref{deSoscel}, except for $q_3 = 0$ and $P_3$ as in Eq. \eqref{lapP3}. {\bf Bottom:} Corresponding filtering at 1000 days.}
\label{deSlapl}
\end{figure}
As mentioned before, the ``real'' Laplace resonance is separated from the de Sitter equilibrium by the rotation and not libration of the $q_3$ argument. It is therefore sufficient to vary the action $P_3$   and therefore the eccentricity of Ganymede (if a fixed semi-major axis is assumed) to distinguish these two. We found that a good value for recovering the main amplitude of the Laplace argument is given by

\begin{equation}\label{lapP3}
P_3 = 10^{-8},
\end{equation}
where the unit of length is assumed normalized with respect to Io's semi-major axis. This value is about two orders of magnitude lower than the value in \citet{CePaPu} (even accounting for the difference in Io's semi-major axis). Furthermore, to avoid being captured in the librational regime, we need to chose the initial value of $q_3$ properly. We here selected $q_3 = 0$.

In Figure \ref{deSlapl} we plot the time evolution of the Hamiltonian Laplace argument, with initial conditions corresponding to the de Sitter equilibrium at second order in the eccentricity, except for $P_3$ and $q_3$, which are taken as above. The results are  clearly greatly improved with respect to Figure \ref{hamLapl}, with the amplitudes of both the normal and filtered angle very close to those in Figure \ref{ephLapl}. The main difference lies in the lack of additional frequencies after filtering, since they have been smoothed out in the Hamiltonian construction. We remark that key to this accurate result is having taken as initial conditions those corresponding to the de Sitter equilibrium at second order in the eccentricity. Using the first-order equilibrium leads to an amplitude of one order of magnitude more for both the normal and the filtered angle.

Interestingly enough, while the diagnostics do vary, the error with respect to the nominal values in Eq. \eqref{laplMM} remains similar to Eq. \eqref{dgnOrHam}. With the initial conditions above, the average over a period of 100 years becomes

\begin{equation}
\begin{aligned}
n_1 - 2n_2 &= 0.733692\ \degree / day,\\
n_2 - 2n_3 &= 0.733676\ \degree / day.
\end{aligned}
\end{equation}
Several factors may contribute to this discrepancy, and it is difficult to pinpoint an exact cause.

To summarize, by exploiting the Hamiltonian dynamics of the system, we have highlighted a way to recover the correct values for the quantities we searched for. However, a quick comparison with the different conditions and results of \citet{CePaPu} shows that this procedure (and therefore the Hamiltonian model) is highly sensitive to the initial conditions. It clearly shows, however, that the approximate nature of the model requires a tradeoff when certain values are searched for. In particular, if the nominal evolution of the Laplace resonance is searched for, the Hamiltonian ``shape'' of the orbits (e.g., semi-major axes and eccentricities) cannot be very close to the real shape.

%
%

\section{Conclusion}

We have considered the problem of constructing approximate models for the dynamics of the Galilean moons in the Jovian system, with a particular focus on recovering the time evolution of the Laplace resonance.
To do this, we considered as benchmark a set of ephemerides extracted from NASA's Spice toolkit, that we defined as elements to recover the corresponding amplitude and period of the main frequency of the Laplace argument, and the related quasi-resonant mean motion relations of the pairs Io-Europa and Europa-Ganymede. Subsequently, we introduced a Jovicentric Cartesian model of the system following \citet{LaThesis}, and exploited it to numerically show that within the scale of 100 years, a sufficient approximation of the Laplace argument can be obtained by considering only the mutual gravitational interactions among the moons, along with Jupiter's oblateness potential truncated at the harmonic $J_2$ (and, of course, including the point-mass gravitational term).

Starting from the information obtained from the Cartesian formulation, we introduced an alternative Hamiltonian model that is more suitable to analytically delve into the mechanisms of the resonance. In order to contain its complexity, we considered the Jovian system as planar, concentrated on the dynamics of the moons involved in the resonance, and restricted the oblateness perturbation to its secular part. 

Numerical comparisons with the Cartesian model showed that if the same initial conditions are used, the Hamiltonian formulation determines the main period of the Laplace argument, but it fails to recover the associated amplitude by an order of magnitude. This is reflected in the mean motion diagnostics, which presents an absolute error on the order of $10^{-3}\ \degree / day$ with respect to the corresponding Cartesian formulation.

By introducing appropriate resonant coordinates, it is apparent that the actual Laplace configuration is one rotating angle away from the only stable equilibrium of the Hamiltonian. Thus, by exploiting normal form techniques to project the Hamiltonian flow onto the appropriate action-angle phase plane, we first numerically located the stable equilibrium, then moved the action variable up to the rotational regime in search of a value of this variable that would generate a good approximation of the amplitude of the Laplace argument. We used a  much lower  value than \citet{CePaPu}. Additionally, the equilibrium was computed using the full Hamiltonian in place of restricting non-oblateness terms to first order in the eccentricity. These two factors combined led to an excellent approximation of the Laplace angle, with an amplitude comparable to the values reported in the literature. Following the chain of canonical transformations backward shows that a good tradeoff to obtain the desired values with this procedure lies in Ganymede's eccentricity.

Parallel and future work involve a thorough use of the Hamiltonian introduced here to determine the mechanisms of multi-body resonances. In particular, \citet{CePaPu-B} considered pairs of commensurabilities different from the two 2:1 studied here, and we classify the dynamics on the base of the eccentricity order associated to the relevant resonant terms. This is a necessary step to study systems different from the Jovian one, which of course means extending the study beyond our solar system. Delving deeper into the mechanisms of these resonances implies extending the time scales considered here, which in turn means considering additional conservative and dissipative perturbations for the models.

%
%

\begin{acknowledgements}
We thank the Italian Space Agency for partially supporting this work through the grant 2013-056-RO.1, and we acknowledge the GNFM/INdAM. We are also grateful to S. Ferraz-Mello for the many fruitful discussions, to L. Iess for his support, and the referee for their useful comments. Finally, F.P. thanks M.J. Mariani, F. De Marchi, and D. Durante for their help with the SPICE toolkit, while A.C. acknowledges the MIUR Excellence Department Project awarded to the Department of Mathematics of the University of Rome ``Tor Vergata'' (CUP E83C18000100006).
\end{acknowledgements}

\bibliographystyle{aa} 
\bibliography{A&A_Draft} 

\begin{appendix}

\section{Explicit Laplace coefficients}

For comparison reasons, in this short appendix we provide an explicit expression for the Hamiltonian terms containing any type of Laplace coefficient (computed through series expansions), that is, the perturbative functions in Eq. \eqref{expHam} and Callisto's contribution, Eq. \eqref{CsHam}. The formulae used for the coefficients can be found in the references mentioned in the appropriate sections, while the values used for the computations are given in the tables of this paper. Of course, the numbers are truncated at a significant enough digit.

The perturbative terms in Eq. \eqref{expHam} are
\begin{equation}
\begin{aligned}
&H_P^{1,2} = -{{Gm_1m_2}\over a_2}\big(0.12954+0.384233e_1\cos(2\lambda_2-\lambda_1-\varpi_1) \\
&-1.18409e_2\cos(2\lambda_2-\lambda_1-\varpi_2) + 0.424985(e_1^2+e_2^2) \\
&+ 1.67945 e_1 e_2\cos(\varpi_1 - \varpi_2) + 3.57298 e_1 e_2 \cos( 4\lambda_2 -2\lambda_1 -\varpi_1 -\varpi_2) \\
&-4.92864 \cos(4\lambda_2 -2\lambda_1 -2\varpi_1) -0.569703 e_2^2 \cos(4\lambda_2 - 2\lambda_1 -2\varpi_2)\big), \\[.1cm]
%
&H_P^{2,3} = -{{Gm_2m_3}\over a_3}\big(0.12861 + 0.379622e_2\cos(2\lambda_3-\lambda_2-\varpi_2) \\
&-1.17533e_3\cos(2\lambda_3-\lambda_2-\varpi_3) + 0.420341(e_2^2+e_3^2) \\
&+ 1.65729 e_2 e_3\cos(\varpi_2 - \varpi_3) + 3.54454 e_2 e_3 \cos( 4\lambda_3 -2\lambda_2 -\varpi_2 -\varpi_3) \\
&- 4.87651 e_2^2 \cos(4\lambda_3 -2\lambda_2 -2\varpi_2) - 0.561562 e_3^2 \cos(4\lambda_3 - 2\lambda_2 -2\varpi_3)\big), \\[.1cm]
%
&H_P^{1,3} = -{{Gm_1m_3}\over a_3}\big(0.04267 + 0.0793481(e_1^2+e_3^2) \\
&+ 0.206309 e_1 e_3\cos(\varpi_1 - \varpi_3) \big),
\end{aligned}
\end{equation}
and Callisto's contribution, Eq. \eqref{CsHam}, can be split into three separate terms (one each for Io, Europa, and Ganymede) as

\begin{equation}
\begin{aligned}
&H_{C}^{1}=-{{Gm_1m_4}\over {a_4}}\ \left[0.01293 + 0.0207567(e_i^2+e_C^2)\right], \\
&H_{C}^{2}=-{{Gm_2m_4}\over {a_4}}\ \left[0.03427 + 0.0615473(e_i^2+e_C^2)\right], \\
&H_{C}^{3}=-{{Gm_3m_4}\over {a_4}}\ \left[0.09993 + 0.253539(e_i^2+e_C^2)\right].
\end{aligned}
\end{equation}

\end{appendix}

\vfill

\end{document}

%% file: Resonance_Geometry.pdf_t
\begin{picture}(0,0)%
\includegraphics{Resonance_Geometry.pdf}
\end{picture}%
\setlength{\unitlength}{4144sp}%
\begingroup\makeatletter\ifx\SetFigFont\undefined%
\gdef\SetFigFont#1#2#3#4#5{%
  \reset@font\fontsize{#1}{#2pt}%
  \fontfamily{#3}\fontseries{#4}\fontshape{#5}%
  \selectfont}%
\fi\endgroup%
\begin{picture}(17679,4973)(2918,-8738)
\put(4951,-8611){\makebox(0,0)[b]{\smash{{\SetFigFont{20}{24.0}{\familydefault}{\mddefault}{\updefault}{\color[rgb]{0,0,0}${\bm 0}$}%
}}}}
\put(9451,-8611){\makebox(0,0)[b]{\smash{{\SetFigFont{20}{24.0}{\familydefault}{\mddefault}{\updefault}{\color[rgb]{0,0,0}${\bm T}$}%
}}}}
\put(13951,-8611){\makebox(0,0)[b]{\smash{{\SetFigFont{20}{24.0}{\familydefault}{\mddefault}{\updefault}{\color[rgb]{0,0,0}${\bm 2}{\bm T}$}%
}}}}
\put(18451,-8611){\makebox(0,0)[b]{\smash{{\SetFigFont{20}{24.0}{\familydefault}{\mddefault}{\updefault}{\color[rgb]{0,0,0}${\bm 3}{\bm T}$}%
}}}}
\end{picture}%

%% file: Prc_Frame.pdf_t
\begin{picture}(0,0)%
\includegraphics{Prc_Frame.pdf}
\end{picture}%
\setlength{\unitlength}{4144sp}%
\begingroup\makeatletter\ifx\SetFigFont\undefined%
\gdef\SetFigFont#1#2#3#4#5{%
  \reset@font\fontsize{#1}{#2pt}%
  \fontfamily{#3}\fontseries{#4}\fontshape{#5}%
  \selectfont}%
\fi\endgroup%
\begin{picture}(6780,7978)(10471,-10601)
\put(12196,-9286){\makebox(0,0)[lb]{\smash{{\SetFigFont{25}{30.0}{\rmdefault}{\mddefault}{\updefault}{\color[rgb]{0,0,0}${\bm \psi}$}%
}}}}
\put(14041,-8566){\makebox(0,0)[lb]{\smash{{\SetFigFont{25}{30.0}{\rmdefault}{\mddefault}{\updefault}{\color[rgb]{0,0,0}${\bm \chi}$}%
}}}}
\put(12061,-5236){\makebox(0,0)[lb]{\smash{{\SetFigFont{25}{30.0}{\rmdefault}{\mddefault}{\updefault}{\color[rgb]{0,0,0}${\bm I}$}%
}}}}
\put(12106,-7351){\makebox(0,0)[lb]{\smash{{\SetFigFont{25}{30.0}{\rmdefault}{\mddefault}{\updefault}{\color[rgb]{0,0,0}${\mathbf P}_0$}%
}}}}
\put(10486,-4786){\makebox(0,0)[lb]{\smash{{\SetFigFont{25}{30.0}{\rmdefault}{\mddefault}{\updefault}{\color[rgb]{0,0,0}$z^\prime$}%
}}}}
\put(12916,-2986){\makebox(0,0)[lb]{\smash{{\SetFigFont{25}{30.0}{\rmdefault}{\mddefault}{\updefault}{\color[rgb]{0,0,0}$z$}%
}}}}
\put(14761,-4426){\makebox(0,0)[lb]{\smash{{\SetFigFont{25}{30.0}{\rmdefault}{\mddefault}{\updefault}{\color[rgb]{0,0,0}$y^\prime$}%
}}}}
\put(17236,-7396){\makebox(0,0)[lb]{\smash{{\SetFigFont{25}{30.0}{\rmdefault}{\mddefault}{\updefault}{\color[rgb]{0,0,0}$y$}%
}}}}
\put(16021,-8611){\makebox(0,0)[lb]{\smash{{\SetFigFont{25}{30.0}{\rmdefault}{\mddefault}{\updefault}{\color[rgb]{0,0,0}$x^\prime$}%
}}}}
\put(10576,-10456){\makebox(0,0)[lb]{\smash{{\SetFigFont{25}{30.0}{\rmdefault}{\mddefault}{\updefault}{\color[rgb]{0,0,0}$x$}%
}}}}
\put(13411,-10456){\makebox(0,0)[lb]{\smash{{\SetFigFont{25}{30.0}{\rmdefault}{\mddefault}{\updefault}{\color[rgb]{0,0,0}$\tilde x$}%
}}}}
\end{picture}%